\documentclass[conference]{IEEEtran}
\IEEEoverridecommandlockouts

\usepackage{cite}
\usepackage{amsmath,amssymb,amsfonts}
\usepackage{algorithmic}
\usepackage{graphicx}
\usepackage{textcomp}
\usepackage{xcolor}
\usepackage{tcolorbox}
\usepackage{booktabs}
\usepackage{ragged2e}

\usepackage{fancybox}
\def\BibTeX{{\rm B\kern-.05em{\sc i\kern-.025em b}\kern-.08em
    T\kern-.1667em\lower.7ex\hbox{E}\kern-.125emX}}
\begin{document}
\pagenumbering{gobble}

\title{Personalized Chain-of-Thought Summarization of Financial News for Investor Decision Support}



\author{\IEEEauthorblockN{1\textsuperscript{st} Tianyi Zhang}
\IEEEauthorblockA{\textit{Computer Science Department} \\
\textit{University of Southern California}\\
Los Angeles, CA, USA \\
tzhang62@usc.edu}
\and
\IEEEauthorblockN{2\textsuperscript{nd} Mu Chen}
\IEEEauthorblockA{\textit{Chief Executive Officer} \\
\textit{Canopy.Cloud}\\
Singapore, Singapore \\
mu.chen@canopy.cloud}
}

\maketitle

\begin{abstract}
Financial advisors and investors struggle with information overload from financial news, where irrelevant content and noise obscure key market signals and hinder timely investment decisions. To address this, we propose a novel Chain-of-Thought (CoT) summarization framework that condenses financial news into concise, event-driven summaries. The framework integrates user-specified keywords to generate personalized outputs, ensuring that only the most relevant contexts are highlighted. These personalized summaries provide an intermediate layer that supports language models in producing investor-focused narratives, bridging the gap between raw news and actionable insights.

\end{abstract}


\section{Introduction}
In today’s financial markets, the stream of information is overwhelming, often repetitive and noisy to investors’ actual needs\cite{Ber2023}. However, extracting relevant insights from this information flow is critical for effective decision-making. Traditional summarization methods, while useful in reducing length, often fail to capture the context \cite{Omar2017} and personalization \cite{Dudy2021} required for financial professionals to act confidently on the information presented.

Recent advances in Large Language Models (LLMs) have created opportunities for more sophisticated text summarization, yet these models often face challenges such as hallucination\cite{Belem2022}, lack of domain grounding\cite{Kang2023}, and limited personalization\cite{Patel2024}. Hallucination represents a critical issue where LLMs generate factually incorrect or fabricated information that appears plausible but contradicts the source material, particularly problematic in financial contexts where accuracy is paramount for investment decisions\cite{Zhang2025}. The lack of domain grounding manifests when models fail to maintain specialized terminology, context-specific relationships, and industry-relevant focus, leading to summaries that may be linguistically coherent but lack the precision required for professional financial analysis\cite{Cao2024}. Limited personalization capabilities prevent these models from adapting their outputs to specific user needs, interests, or expertise levels, resulting in generic summaries that may not highlight the most relevant information for particular stakeholders such as financial advisors focusing on specific market sectors or investment strategies\cite{Kirs2024}.

These shortcomings make it difficult for financial professionals to fully trust or act on automatically generated outputs, as inaccurate information can lead to significant financial losses and regulatory compliance issues. To address these challenges, Chain-of-Thought (CoT) reasoning has emerged as a promising approach~\cite{Li2024}~\cite{Lee2025}. By explicitly modeling intermediate reasoning steps, CoT methods improve coherence, logical flow, and interpretability, enabling summaries that filter out irrelevant details while maintaining structured and contextually faithful content~\cite{Chen2025}. However, existing CoT-based summarization approaches remain largely generic, lacking mechanisms to adapt to the specific priorities and information needs of individual investors or advisors~\cite{Iz2025}~\cite{Zhang2024PersonalSum}. Most current approaches employ uniform reasoning chains that do not differentiate between various types of financial information, treating earnings reports, merger announcements, and regulatory updates with the same generic processing framework~\cite{Qzhang2020}. Furthermore, these methods lack domain-specific knowledge integration, failing to recognize the hierarchical importance of quantitative metrics such as revenue growth, profit margins, and valuation multiples that are critical for investment decision-making~\cite{SS2021}. The reasoning chains in existing CoT approaches also do not incorporate user-specific context, such as investment focus areas and risk tolerance, which are essential for generating actionable financial insights~\cite{Vansh2023}. This generic approach results in summaries that, while logically coherent, fail to provide the targeted, quantitative, and contextually relevant information that financial professionals require for effective decision-making.

To address this gap, we propose a personalized CoT summarization framework for financial news. Our approach generates concise, event-driven summaries that are tailored to user-defined keywords, ensuring that only the most relevant information are highlighted. These keyword-driven summaries function as an intermediate representation, allowing LLMs to produce narratives that are both contextually rich and directly aligned with investor interests. In doing so, the framework bridges the gap between raw financial reporting and actionable investor insights.

Our experimental evaluation demonstrates the effectiveness of this approach: our enhanced summaries achieve a BLEU score of 0.1786 and ROUGE-L score of 0.4028, representing substantial improvements of 267\% and 90\% respectively over GPT-4o generated summaries (Table~\ref{tab:evaluation_results}). Furthermore, our personalized keyword-based binary relevance classification outperforms commonly used ranking selection methods by 40\% in accuracy (Table~\ref{tab:classification_results}), demonstrating superior precision in identifying articles relevant to specific investor interests. These results validate that our specialized CoT framework not only produces higher-quality financial summaries but also delivers more accurate personalization for investment decision support.

The contributions of this paper are twofold:

\begin{enumerate}
    \item \textbf{Framework Design} -- We introduce a CoT-based summarization pipeline that condenses financial news into concise, event-focused outputs.
    
    \item \textbf{Personalization Mechanism} -- We incorporate keyword-driven filtering to adapt summaries to the unique needs of financial advisors and investors.
    
\end{enumerate}

\section{Related Work}

\subsection{Financial News Summarization and Domain Adaptation}
Financial news summarization (FNS) has been studied as a way to digest market-moving information efficiently. Summaries of financial reports and news are expected to retain events and signals that could affect market behavior~\cite{Mou2025newsum}. Early approaches ranged from extractive techniques, such as selecting and ranking key sentences to fully abstractive generation. Graph-based methods such as TextRank have been applied to extract key sentences from stock news~\cite{Mihalcea2004}. While extractive methods gather critical information, they struggle with long-range dependencies and offer no personalization. Abstractive summarization with encoder--decoder models such as BART or PEGASUS can produce more coherent financial summaries~\cite{Zhang2020PEGASUS}, but these models risk hallucinating unsupported details and require large domain-specific datasets, which are costly to create. 

Domain adaptation has proven effective in addressing these challenges. FinBERT, for example, is pre-trained on financial corpora to capture the field’s unique context~\cite{Liu2020FinBERT}. However, it is important to note that FinBERT is not designed for summarization; rather, it functions as a domain-adapted encoder for classification or embedding tasks. This highlights effective summarization architectures are still required to produce coherent and relevant summaries. Prior work has also explored financial summarization for investment insights, e.g., generating news summaries to assist stock prediction~\cite{Ding2016}, or company-focused news headlines~\cite{Feng2022}. 

Beyond task-specific models, broader surveys and evaluations have examined the intersection of NLP and finance. Du et al. provide a comprehensive survey of NLP in finance, covering applications such as sentiment analysis, narrative processing, forecasting, and summarization, while also outlining persistent challenges related to domain adaptation, data quality, and specialized summarization~\cite{Du2025NLPFinance}. Complementing this, Du et al. assess the reasoning capabilities of LLMs in financial sentiment analysis, showing that even advanced models like GPT-3.5 and PaLM-2 are particularly weak in numerical and comparative reasoning~\cite{Du2025LLMReasoning}. Similarly, Cambria et al. discuss the opportunities and challenges of applying LLMs in financial contexts, noting that issues of factuality, domain grounding, and personalization remain unresolved for decision-support applications~\cite{Cambria2024LLMFinance}. Nonetheless, most existing financial summarization systems still produce one-size-fits-all outputs and lack mechanisms to adapt to an individual investor’s focus. Our work addresses this gap by combining domain-aware summarization with personalized filtering.

\subsection{Personalized and Controllable Summarization}
Personalized summarization aims to tailor outputs to the reader’s needs and remains an under-explored area~\cite{Ao2021, Dudy2021}. Recent LLMs achieve strong performance on generic news summarization~\cite{Zhang2020PEGASUS}, yet generic summaries often fail to meet specific user intentions~\cite{Zhang2024PersonalSum}. Traditional datasets rarely capture user preferences, limiting research to generic summary generation. In response, controllable summarization methods were introduced, where models are guided with attributes such as length or topic~\cite{He2020CTRLsum, Dou2021GSum}. These methods demonstrate guided outputs are feasible, but their controls are usually generic rather than user-specific. Ao et al.~\cite{Ao2021} introduced PENS, a dataset for personalized news headline generation, while Zhang et al.~\cite{Zhang2024PersonalSum} proposed PersonalSum, designed to evaluate user-tailored summaries. These works confirm that while personalization is desirable, most current summarizers still output the same summary for everyone. In finance, this gap is pronounced: wealth advisors and retail investors have distinct information needs, but existing systems cannot tailor summaries accordingly. Typically, personalization has been addressed by recommendation systems (filtering articles) rather than adapting summary content itself.

\subsection{Reasoning and Chain-of-Thought in Summarization}
Chain-of-thought (CoT) prompting has demonstrated that guiding LLMs to produce intermediate reasoning steps can improve performance on complex tasks~\cite{Wei2022}. Instead of producing an answer directly, models generate a reasoning process, yielding state-of-the-art results on arithmetic, commonsense, and symbolic reasoning benchmarks~\cite{Wei2022}. Although summarization is not a classical reasoning problem, financial news requires reasoning about event importance and user relevance. Traditional summarizers often operate in an end-to-end fashion, missing implicit but crucial signals. Zhu et al.~\cite{Zhu2025} note that standard CoT prompting may fail when essential information is implicit, since the model may overlook early extraction of key facts. They propose iterative summarization pre-prompts to distill relevant content before reasoning, which improved accuracy. This aligns with our strategy of using keyword-filtered mini-summaries as intermediate representations. Multi-stage summarization methods, such as extract-then-abstract pipelines~\cite{Nallapati2017} or recursive summarization frameworks~\cite{Cao2015}, similarly use intermediate steps to improve coverage, though they are not personalized. Our framework extends this line of work by integrating user-defined keyword filtering into the reasoning chain, ensuring the final summary is both contextually rich and personalized to investor needs.

\section{Methodology}
\begin{figure*}[htbp]
\centerline{\includegraphics[width=1\textwidth]{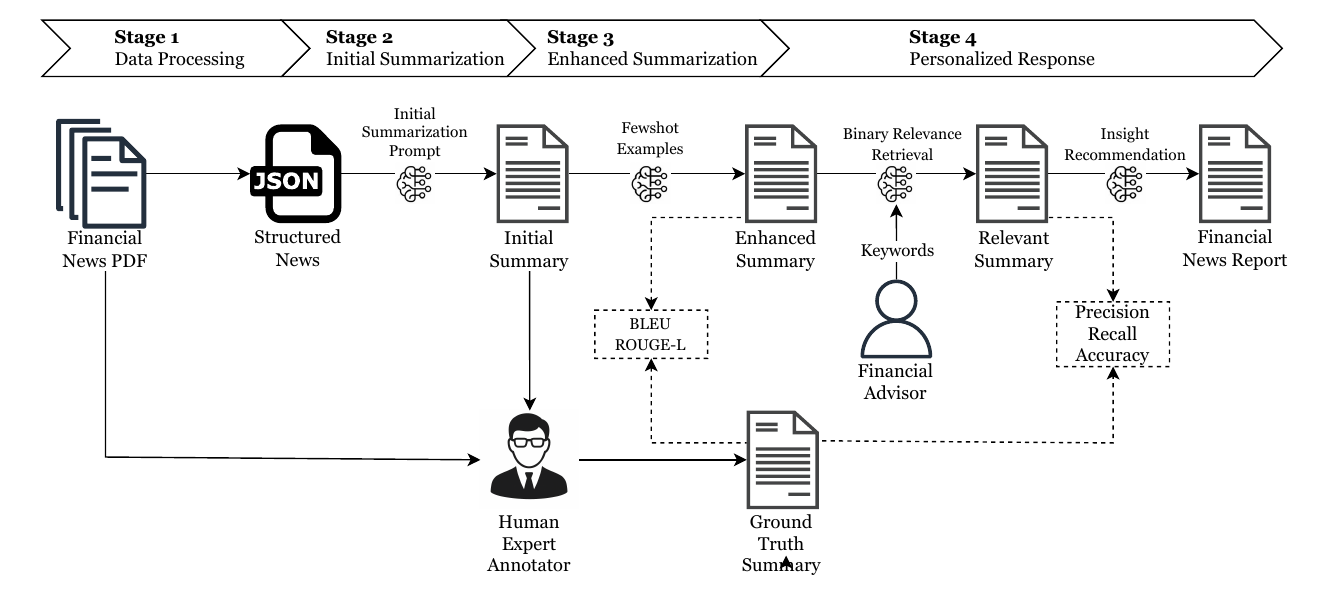}}
\caption{Overview of Personalized Chain-of-Thought Financial News Summarization Framework}
\label{figure:framework}
\end{figure*}

Our proposed methodology implements a personalized chain-of-thought summarization pipeline for financial news analysis, designed to support investor decision-making. As shown in Fig~\ref{figure:framework}, the system operates through four sequential stages, each leveraging LLMs to progressively refine and contextualize financial information.

\begin{enumerate}
    \item \textbf{Data Processing Module}: PDF text extraction and English language filtering
    \item \textbf{Initial Summarization Module}: First-pass article summarization for financial investors
    \item \textbf{Enhanced Summary Processing Module}: Metadata extraction and financial-focused summary refinement
    \item \textbf{Personalized Response Generation Module}: Keyword-based relevance assessment and investment recommendations
\end{enumerate}

\subsection{Data Processing and Text Extraction}

The first stage addresses the challenge of extracting structured text from diverse financial news PDFs. We implement a dual-approach extraction strategy using both PyPDF2\cite{pypdf2} and pdfplumber\cite{pdfplumber} libraries to ensure robust text extraction across different PDF formats and layouts.

\subsubsection{Text Extraction Pipeline}

For each PDF document $D_i$ in the input directory, we apply a configurable extraction strategy where $T_i = \text{Extract}(D_i)$ represents the extracted text from document $D_i$. Our framework defaults to pdfplumber extraction due to its superior handling of complex layouts commonly found in financial news publications, with PyPDF2 available as an alternative for simpler document structures. This dual-library approach ensures robust text extraction capabilities across diverse PDF formats while prioritizing the more reliable pdfplumber method for financial document processing.

\subsubsection{English Language Filtering}

We implement a language filtering mechanism that retains only English-dominated content. Since the financial news articles in our dataset contain bilingual content (English and Chinese), we apply sentence-level filtering to extract English portions. For each sentence $s_j$ in extracted text $T_i$, we calculate the English character ratio:

\begin{equation}
R_{eng}(s_j) = \frac{|\text{EnglishChars}(s_j)|}{|s_j|}
\end{equation}

Sentences are retained if $R_{eng}(s_j) > 0.7$, ensuring that financial terminology and market-specific language are preserved while filtering out Chinese content.

\subsubsection{Metadata Preservation}

Each processed document maintains comprehensive metadata including: (a) source file information, (b) extraction method used, (c) file size and text length statistics, and (d) language filtering status.

 \subsection{Initial Financial Summarization}

\begin{figure}[htbp]
\centerline{\includegraphics[width=0.48\textwidth]{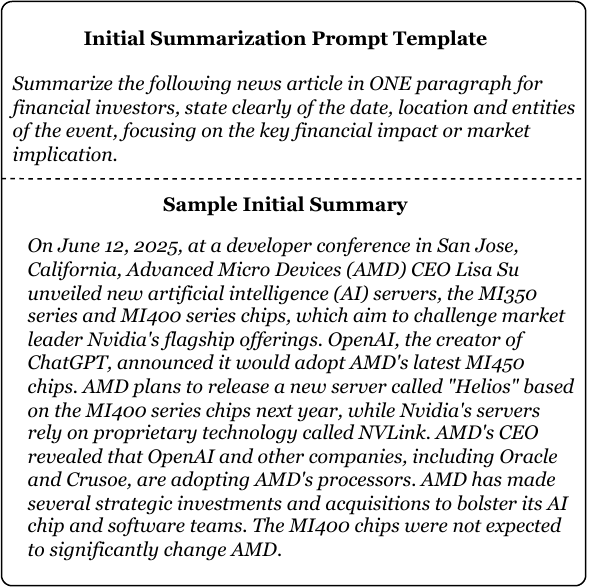}}
\caption{Second Stage: Initial Summarization Prompt Template and Sample Output}
\label{fig:initial_summarization}
\end{figure}

 The second stage employs the Mistral-7B-Instruct-v0.2 model~\cite{mistral} to generate initial financial-focused summaries, implementing the first layer of chain-of-thought reasoning by transforming raw text into structured financial narratives. We design specialized prompts that guide the LLM to focus on financial implications, with the complete prompt template and example results shown in Figure~\ref{fig:initial_summarization}. The summarization process uses controlled generation parameters including a maximum of 200 tokens for concise financial summaries and a temperature of 0.7 to balance creativity and consistency in the generated outputs.

\subsection{Stage 3: Enhanced Summary Processing with Few-Shot Learning}

The third stage implements advanced metadata extraction and summary refinement through few-shot learning, representing a critical enhancement in our chain-of-thought approach. We leverage professional financial analyst expertise to guide the LLM's output quality and focus.

\subsubsection{LLM-Based Metadata Extraction}

We employ the same Mistral model to extract structured metadata from article content:

\begin{equation}
M_i = \text{ExtractMetadata}(T_i) = \{\text{date}, \text{location}, \text{entity}\}
\end{equation}

The metadata extraction uses a JSON-structured prompt that forces the model to output parseable structured data, with fallback mechanisms for cases where LLM extraction fails.

\subsubsection{Few-Shot Learning for Financial Summary Enhancement}

The enhanced summary generation implements a sophisticated few-shot learning approach using professional financial analyst examples. We provide the LLM with curated examples that demonstrate the desired output format and analytical depth as shown in Figure~\ref{fig:fewshot}.

\begin{figure}[htbp]
\centerline{\includegraphics[width=0.49\textwidth]{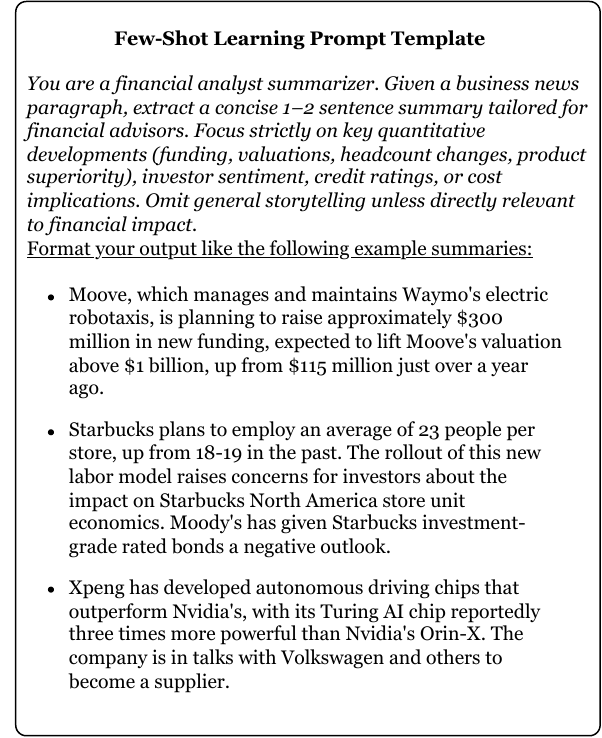}}
\caption{Third Stage: Enhanced Summary Few-Shot Learning Prompt Template}
\label{fig:fewshot}
\end{figure}

\subsection{Stage 4: Personalized Response Generation}

The final stage implements keyword-based personalization and investment action synthesis. Financial advisors input specific keywords of interest (e.g., ``AI", ``United States", ``new energy") to retrieve personalized investment insights tailored to their focus areas.

\subsubsection{Relevance Assessment Pipeline}

For a given keyword $K$ provided by the financial advisor, we evaluate each article $A_i$ using a Binary Relevance Retrieval approach:

\begin{equation}
R(A_i, K) = \text{LLM\_Classify}(A_i, K) \in \{\text{YES}, \text{NO}\}
\end{equation}

The relevance assessment considers direct sector connections, entity relationships, and market impact relevance to the specified keyword.

\subsubsection{Key Insights Generation}

Based on relevant articles $\{A_r\}$, we generate key insights $I$:

\begin{equation}
I = \text{GenerateInsights}(K, \{A_r\})
\end{equation}

Each insight follows a structured format that identifies trends, financial implications, and risk/opportunity factors specific to the advisor's keyword interest.

\subsubsection{Investment Action Synthesis}

The final step synthesizes specific investment actions $A_{inv}$:

\begin{equation}
A_{inv} = \text{GenerateActions}(K, \{A_r\}, I)
\end{equation}

Investment actions are constrained to exactly three actionable recommendations, each focusing on specific opportunities or risk management strategies relevant to the financial advisor's specified keyword. A sample keyword input and the corresponding final personalized report are illustrated in Figure\ref{fig:fewshot}.

\begin{figure}[htbp]
\centerline{\includegraphics[width=0.49\textwidth]{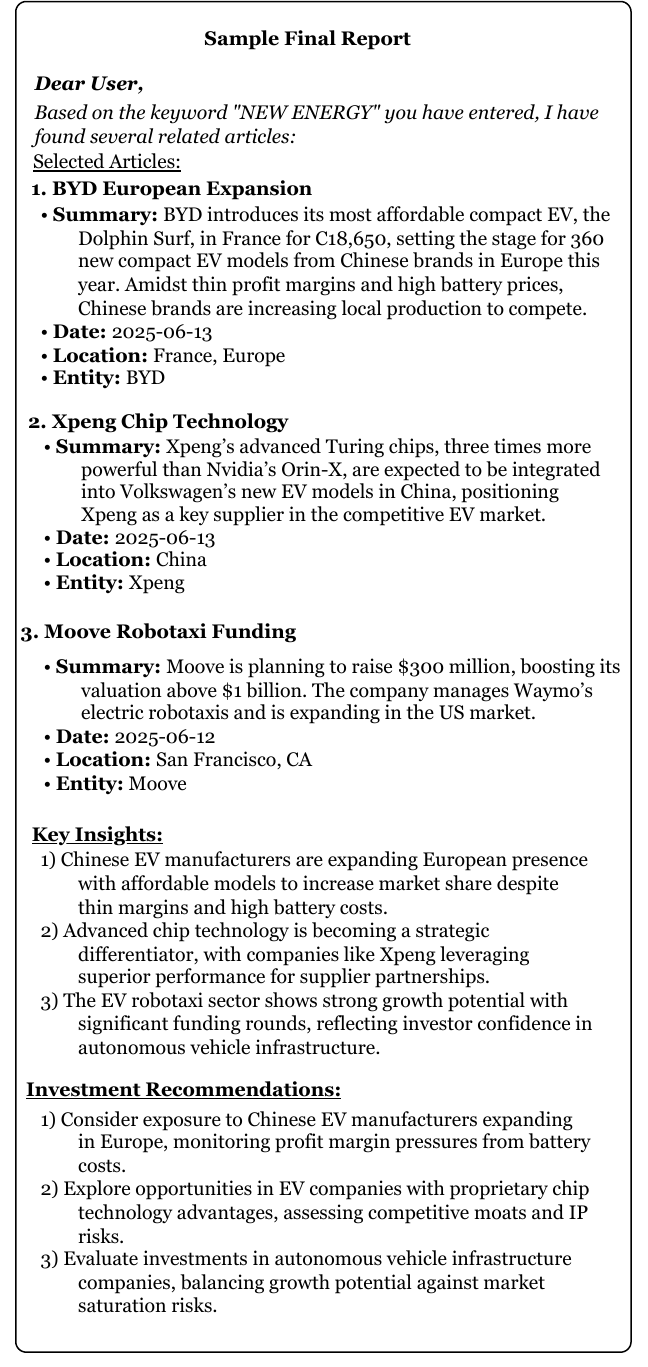}}
\caption{Fourth Stage: Personalized Response Generation Sample Output}
\label{fig:fewshot}
\end{figure}

\subsection{Chain-of-Thought Reasoning Implementation}

Our methodology implements chain-of-thought reasoning through sequential LLM interactions that build upon previous outputs:

\begin{equation}
\text{Chain}_i = f(\text{Chain}_{i-1}, \text{Input}_i, \text{Context}_i)
\end{equation}

where each stage $i$ uses the output from stage $i-1$ as context for the next reasoning step. The few-shot learning in Stage 3 ensures that the enhanced summaries maintain professional financial analyst standards, improving the quality of downstream analysis in Stage 4.

\subsection{Model Configuration and Optimization}

Throughout all stages, we maintain consistent model parameters to ensure reproducibility and coherence across the pipeline. We employ Mistral-7B-Instruct-v0.2 as the base model for all text generation tasks, leveraging automatic GPU allocation for optimal computational efficiency. The model operates with a context length of 2048 tokens with truncation applied when input exceeds this limit.

\section{Evaluation}
\subsection{Evaluation of Enhanced Summary}

We evaluate our personalized chain-of-thought summarization framework using both automated metrics and professional financial analyst assessment. The evaluation compares our enhanced summaries (Stage 3 output) against ChatGPT-generated summaries using the same ground truth references.
\subsubsection{Dataset Creation}

To establish reliable ground truth references for evaluation, a professional financial analyst with over 10 years working experience created high-quality reference summaries. The ground truth generation process involved the analyst independently reviewing the articles and initial summaries produced in the second stage, then manually identifying the most important sentences and financial information. The highlighted information was then synthesized into concise ground truth summaries that capture the essential financial insights. This human-curated ground truth dataset serves as the reference standard for calculating BLEU and ROUGE-L scores, ensuring that our automated evaluation metrics reflect professional financial analysis standards while maintaining focus on quantitative developments and market implications essential for investment decision-making.

\subsubsection{Evaluation Results}

Table~\ref{tab:evaluation_results} presents the comparative evaluation results between our enhanced summaries and GPT-4o generated summaries across seven financial news articles.

\begin{table}[htbp]
\centering
\caption{Automated Evaluation Results: Average BLEU and ROUGE-L Scores}
\label{tab:evaluation_results}
\begin{tabular}{lcc}
\toprule
\textbf{Summary Type} & \textbf{BLEU Score} & \textbf{ROUGE-L Score} \\
\midrule
Enhanced Summary (Ours) & \textbf{0.1786} & \textbf{0.4028} \\
GPT-4o Summary      & 0.0487          & 0.2123 \\
\midrule
\textbf{Improvement}    & \textbf{+267\%} & \textbf{+90\%} \\
\bottomrule
\end{tabular}
\end{table}

\subsubsection{Performance Analysis}

Our enhanced summarization framework demonstrates superior performance across both evaluation metrics:

\begin{itemize}
    \item \textbf{BLEU Score}: Our enhanced summaries achieve an average BLEU score of \textbf{0.1786}, representing a \textbf{267\%} improvement over ChatGPT summaries (0.0487).
    \item \textbf{ROUGE-L Score}: Our approach yields an average ROUGE-L score of \textbf{0.4028}, outperforming ChatGPT summaries (0.2123) by \textbf{90\%}.
\end{itemize}

These results demonstrate that our few-shot learning approach with professional financial analyst examples significantly enhances summary quality and alignment with ground truth references.

\subsubsection{Professional Financial Analyst Assessment}

To complement our automated evaluation, we conducted a qualitative assessment with professional financial analysts from leading investment firms. Two certified financial analysts independently evaluated both summary types across five key criteria: financial relevance, professional quality, actionability, conciseness, and factual accuracy. Both analysts consistently preferred our enhanced summaries over GPT-4o generated summaries across all evaluation criteria, noting that our summaries demonstrated superior focus on quantitative developments such as funding amounts, valuation changes, and market share data, while maintaining professional terminology and financial context more accurately. 

\begin{figure}[htbp]
\centerline{\includegraphics[width=0.49\textwidth]{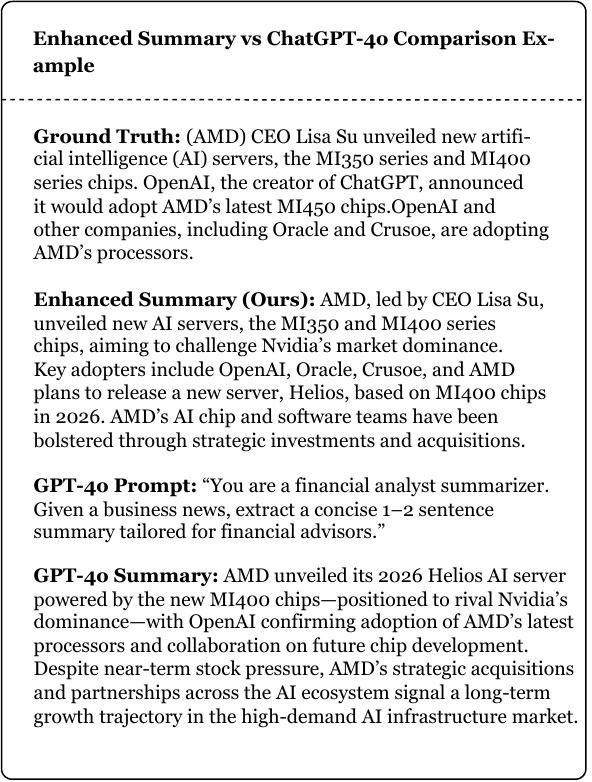}}
\caption{Sample Summary Comparison}
\label{fig:comparison}
\end{figure}

The analysts highlighted that enhanced summaries articulated investment implications more clearly, aligned better with industry reporting standards, and critically maintained factual accuracy without hallucinations, ensuring that all financial concepts, metrics, and market references remained anchored to the original source material without fabrication. This professional validation confirms that our chain-of-thought approach with few-shot learning effectively bridges the gap between automated summarization and professional financial analysis standards while maintaining critical factual integrity essential for financial decision-making. A sample comparison between enhanced summary and GPT-4o summary is illustrated in Figure~\ref{fig:comparison}.

\subsection{Personalization Evaluation: Selection Strategy Comparison}

To evaluate the effectiveness of our personalization framework (Stage 4), we compare two approaches for keyword-based article selection: binary relevance classification versus multi-article ranking selection. This evaluation examines how different selection strategies affect the quality of article retrieval for personalized investment insights.

\subsubsection{Selection Approaches}

We implement two distinct selection methodologies:

\begin{itemize}
    \item \textbf{Binary Relevance Retrieval}: Each article is independently evaluated for relevance to the keyword, with the LLM outputting a binary decision (1 = relevant, 0 = not relevant) for each article
    \item \textbf{Multi-Article Ranking Selection}: The LLM simultaneously evaluates all articles and selects the most relevant subset from the entire collection in a single decision process
\end{itemize}

\subsubsection{Evaluation Methodology}

Professional financial analysts manually annotated articles for relevance to various financial keywords (e.g., "AI", "China", "new energy", "funding"). We then compared both LLM selection approaches against these human annotations using accuracy as the primary metric.

\subsubsection{Selection Performance Results}

Table~\ref{tab:classification_results} presents the comparative performance between Binary Relevance Retrieval and multi-article ranking selection approaches for keyword relevance assessment.

\begin{table}[htbp]
\centering
\caption{Article Selection Approach Comparison: Accuracy Results}
\label{tab:classification_results}
\begin{tabular}{lccc}
\toprule
\textbf{Selection Strategy} & \textbf{Accuracy} & \textbf{Precision} & \textbf{Recall} \\
\midrule
Binary Relevance Retrieval      & \textbf{0.8750} & \textbf{0.8421} & \textbf{0.8889} \\
Multi-Article Ranking Selection & 0.6250 & 0.6154 & 0.6667 \\
\midrule
\textbf{Improvement}        & \textbf{+40\%}  & \textbf{+37\%}  & \textbf{+33\%} \\
\bottomrule
\end{tabular}
\end{table}

\subsubsection{Analysis of Selection Results}

The Binary Relevance Retrieval approach demonstrates significantly superior performance across all evaluation metrics. The binary method achieves an accuracy of \textbf{87.5\%} compared to 62.5\% for multi-article ranking selection, representing a \textbf{40\% improvement}. 

Based on these results, our framework adopts the Binary Relevance Retrieval approach for Stage 4 personalization, ensuring optimal relevance assessment for keyword-based article selection. This design choice directly contributes to the quality of personalized investment insights by ensuring that only the most relevant articles are selected for further analysis and recommendation generation.

\section{Conclusion}

By combining structured reasoning with personalization, our framework provides a scalable solution to managing financial news overload. The work advances the use of LLMs in financial text analysis. More importantly, it contributes to broader efforts in applying AI to decision-support systems in dynamic and information-rich environments.

Our experimental validation demonstrates the practical effectiveness of this approach, with enhanced summaries achieving substantial improvements of 267\% in BLEU scores and 90\% in ROUGE-L scores compared to GPT-4o, while our binary classification strategy for personalization outperforms traditional ranking methods by 40\% in accuracy. These quantitative improvements, coupled with consistent professional analyst preference for our enhanced summaries, validate the framework's ability to bridge the gap between automated text processing and professional financial analysis standards.

The implications of this work extend beyond financial news summarization. The chain-of-thought approach with few-shot learning demonstrates how domain-specific expertise can be effectively integrated into large language model workflows, providing a template for similar applications in healthcare, legal analysis, and other specialized fields where accuracy and professional standards are paramount. The personalization mechanism offers a scalable approach to tailoring AI-generated content to individual user needs without requiring extensive fine-tuning or domain-specific model training.

Looking forward, this framework establishes a foundation for more sophisticated financial AI systems that could integrate real-time market data, regulatory updates, and individual investor profiles to provide comprehensive decision support. As financial markets become increasingly complex and data-driven, such AI-assisted analysis tools will become essential for maintaining competitive advantage while managing information complexity. The success of our approach suggests that the future of financial AI lies not in replacing human expertise, but in augmenting professional capabilities through intelligent, personalized information processing systems.

\section{Limitations and Future Work}

While our framework successfully generates concise, information-rich summaries tailored for financial advisors and investors, several limitations warrant discussion. Although the enhanced summaries effectively highlight key quantitative information and market implications, current large language models lack the sophisticated financial reasoning capabilities required to provide definitive investment guidance. Financial decision-making involves complex considerations including risk tolerance, portfolio diversification, market timing, regulatory factors, and individual investor circumstances that extend beyond the scope of news summarization. The generated investment recommendations should therefore be viewed as preliminary insights that require further analysis and validation by qualified financial professionals rather than actionable trading decisions.

Additionally, our framework employs a four-stage pipeline that could potentially be simplified for certain applications. For instance, Stage 2 (initial summarization) and Stage 3 (enhanced processing with few-shot learning) could theoretically be combined into a single step to reduce computational overhead and processing time. However, we deliberately separated these stages to facilitate the creation of reliable ground truth references, as professional financial analysts found it more intuitive to evaluate and refine initial summaries rather than working directly with raw article text. This staged approach enabled more faster human annotation and quality assessment, though future research could explore more streamlined architectures that maintain summary quality while reducing system complexity.

Future work should investigate the integration of domain-specific financial knowledge bases, real-time market data, and personalized risk assessment models to enhance the framework's decision-support capabilities. Additionally, research into more sophisticated prompt engineering techniques and fine-tuning approaches could potentially consolidate the multi-stage pipeline while preserving the quality benefits of our current methodology.




\end{document}